\documentclass[prc,preprint]{revtex4}
\usepackage{graphicx}
\usepackage{amsmath,amssymb}
\usepackage{mathrsfs}
\def\pt{p_T}
\def\v{$v_2^h(\pt,b)$}
\def\dis{distribution}
\def\pb{\phi,b}
\def\ppb{p_T,\phi,b}
\def\bq{\begin{eqnarray}}
\def\eq{\end{eqnarray}}
\begin{document}

\title{\Large {\bf Effects of Minijets on Hadronic Spectra and Azimuthal Harmonics in Au-Au Collisions at 200 GeV}}
\author
 {\bf Rudolph C. Hwa$^1$ and Lilin Zhu$^2$}
\affiliation
{$^1$Institute of Theoretical Science and Department of
Physics\\ University of Oregon, Eugene, OR 97403-5203, USA\\
$^2$Department of  Physics, Sichuan
University, Chengdu  610064, P.\ R.\ China}
\date{\today}

\begin{abstract}
{The production of hadrons in heavy-ion collisions at RHIC in the low transverse-momentum ($p_T$) region is investigated in the recombination model with emphasis on the effects of minijets on the azimuthal anisotropy. Since the study is mainly on the hadronization of partons at late time, the fluid picture is not used to trace the evolution of the system.
The inclusive distributions at low $p_T$ are determined as the recombination products of thermal partons. The $p_T$ dependencies of both pion and proton have a common exponential factor apart from other dissimilar kinematic and resonance factors, because they are inherited from the same pool of thermal partons. Instead of the usual description based on hydrodynamics, the azimuthal anisotropy of the produced hadrons is explained as the consequence of the effects of minijets, either indirectly through the recombination of enhanced thermal partons in the vicinity of the trajectories of the semihard partons, or directly through thermal-shower recombination.
Although our investigation is focussed on the single-particle distribution at midrapidity, we give reasons why a component in that distribution can be identified with the ridge, which together with the second harmonic $v_2$ is due to the semihard partons created near the medium surface that lead to calculable anisotropy in $\phi$.
It is shown that the higher azimuthal harmonics, $v_n$, can also be well reproduced without reference to flow.  The $p_T$ and centrality dependencies of the higher harmonics are prescribed by the interplay between TT and TS recombination components. The implication of the success of this drastic departure from the conventional approach is discussed.}

\end{abstract}
\maketitle

\section{Introduction}
As the data on single-particle distributions of identified hadrons produced in heavy-ion collisions become more abundant and precise \cite{sa, ja, ba, sa1, igb, iga, ba1}, more demands are put on theoretical models to reproduce them.
It is generally recognized that in Au-Au collisions at $\sqrt{s_{NN}}=200$ GeV at the Relativistic Heavy-Ion Collider (RHIC) the low transverse-momentum ($\pt$) region ($\pt<2$  GeV/c) is well described by hydrodynamics \cite{es,uh,gm,rr,dat}. Because it has attained the status of the conventional approach, it is of interest to point out that an alternative approach without using the fluid description can also reproduce the same data on $\pt$ and azimuthal angular ($\phi$) dependencies and without using more parameters. The approach that we present here does not have the virtue of tracking the time evolution of the dense system, but it presents a different point of view on the origin of the azimuthal asymmetry. Instead of emphasizing early thermalization and the effects of anisotropic pressure gradient, we consider the non-flow effects of minijets due to semihard scattering  of partons. The basic idea has been discussed previously \cite{rch,chy,hz}. Here we present a more detailed phenomenological analysis of the $\pt$ and $\phi$ distributions of pions and protons produced at RHIC for a range of centralities.

Our calculations are based on the recombination model \cite{hy}, which is a particular implementation of the general approach of coalescence  that has been shown to be successful in the intermediate region, $2<\pt<6$ GeV/c \cite{hy1, gkl, fmnb}. Here we push to the lower region $\pt<2$ GeV/c, in which the thermal partons are dominant. However, because semihard partons can lose energy to the thermal medium and result in local enhancement that is azimuthally anisotropic, there are non-trivial complications in the thermal sector.
The point that motivates our study is related to the question of what happens to the initial system within 1 fm/c after collision.  Semihard partons created within 1 fm from the surface will have already left the initial overlap region before thermalization is complete.   They are the minijets that can give rise to $\phi$ dependence, not accounted for by conventional hydrodynamics.   When the parton $k_T$ is low enough so that minijets are copiously produced, the corresponding effect on the $\phi$ anisotropy can become dominant, and is insensitive to the type of hadron produced.

Another area of concern is the  $p_T$ \dis s in the low-$p_T$ region, where pion and proton appear empirically to have different behaviors.  In the parton recombination model the hadrons should have the same inverse slope as that of the coalescing quarks if the hadrons are formed by recombination of the thermal partons, but because of the difference in the meson and baryon wave functions, the net $p_T$ distributions turn out to be different.  This line of analysis takes into account the quark degree of freedom just before hadronization, which is overlooked by the fluid description of the flow effect.  The burden is to show that the data on $v_2(p_T)$ can be reproduced for both pion and proton at low $p_T$  in an approach using a common inverse slope $T$ without relying explicitly on the hydro description of elliptic flow.

The basic assertion in our study is that the recombination of thermal partons has two components, one is azimuthally isotropic, called the Base, while the other one is identified with the Ridge, which has the $\phi$ dependence that is calculable. Our focus is on single-particle \dis s at midrapidity and low $\pt$ and on how they are affected by minijets. The semihard partons that give rise to the observed minijets generate also the second component in the inclusive \dis. It will be our main task to show that the second component exhibits the properties of both the ridge and the second harmonic in $\phi$.

Recently there are experimental and theoretical studies of higher harmonic coefficients, $v_n$,  of the azimuthal asymmetry that have been related to the fluctuations of the initial configuration of the collision system \cite{ar,ty,xk,blo,ps,ral}. Although the phenomena are not of first order in importance compared to the second harmonics $v_2$, which has been regarded as the primary evidence for hydrodynamical flow \cite{kh}, it is imperative  for us to explain their origin in our approach that has no explicit formalism to connect the initial and final states. We shall show that their dependence on $\pt$ and centrality can be well accounted for by the thermal-shower component of recombination, so the minijets themselves cause the $\phi$ anisotropy that leads to $v_n$, whereas the dominant phenomenon in $v_2$ is due to the enhancement in the thermal-thermal sector.

We are aware that our approach is not in the mainstream and that we do not have a code to simulate the evolution of the dense system. However, it is of some value to have explicit analytic expressions that show why the pion and proton \dis s in $\pt$ and $\phi$ have common properties based on the parton \dis s before hadronization, and how minijets can affect azimuthal harmonics  in ways that are in accord with the measured behaviors. To have phenomenological evidence for the validity of an alternative approach that does not rely on hydrodynamical flow enriches the scope of inquiry into the various processes that can be important in heavy-ion collisions and may even cast doubt on the validity of the assumptions made in the conventional approach.

 Before entering into the details of our formalism, it is helpful to clarify  possible confusions of what we do with  the conventional approach when common terms are used with different meanings. In hydrodynamical treatment of the dense system, it is usually assumed that rapid thermalization is completed in less than 1 fm/c and that the expansion of the system can then be described by the hydro equations with suitable assumptions about equations of state and viscosity. Thus the words hydro and thermal are almost synonymous. In our treatment we use thermal without implying hydro. That is because we apply the notion of thermal to the soft partons only at late time just before hadronization. If the system takes over 5 fm/c to equilibrate fully, that would invalidate the use of hydrodynamics from early time, but would not affect the validity of our approach. Or, if minijets  make enhanced thermal contribution to the soft sector through energy loss of semihard partons to the medium without being a part of the equilibrated system from the beginning, then the thermal sector describable by hydro consists of only a portion of the soft hadrons in the final state, leaving another soft (but thermal) component that is outside hydro. These are possibilities that do not invalidate our approach; indeed, our treatment is aimed at accounting for the effects of those minijets.

\section{Common Form of Hadronic Spectra}

We begin with a recapitulation of our description of single-particle distribution \cite{hy}.  At low $p_T$ we consider first the recombination of thermal partons, for which the pion and proton spectra at $y=0$ are given by
\begin{eqnarray}
p^0 {dN^{\rm TT}_{\pi}\over dp_T} &=& \int \prod_{i=1}^2 \left[{dq_i\over q_i} {\cal T} (q_i)\right] {\cal R}_{\pi}(q_1,q_2,p_T),     \label{1} \\
p^0 {dN^{\rm TTT}_p\over dp_T} &=& \int \prod_{i=1}^3 \left[{dq_i\over q_i} {\cal T} (q_i)\right] {\cal R}_p(q_1,q_2,q_3,p_T),     \label{2}
\end{eqnarray}
where ${\cal T} (q_i)$ is the thermal distribution of the quark (or antiquark) with momentum $q_i$, and ${\cal R}_h$ is the recombination function (RF) for $h=\pi$ or $p$.  On the assumption that collinear quarks make the dominant contribution to the coalescence process (so that the integrals are one-dimensional for each quark along the direction of the hadron), the RFs are
\begin{eqnarray}
{\cal R}_{\pi}(q_1,q_2,\pt) &=& {q_1q_2\over \pt^2} \delta\left(\sum_{i=1}^2 {q_i\over \pt} - 1 \right),    \label{3} \\
{\cal R}_p(q_1,q_2,q_3,\pt) &=& f_p\left({q_1\over p_T},{q_2\over \pt},{q_3\over \pt} \right) \delta\left(\sum_{i=1}^3 {q_i\over \pt} - 1 \right)     \label{4}
\end{eqnarray}
where the details of $f_p(q_i/p_T)$ that depends on the proton wave function are given in \cite{hy}, and need not be repeated here.  The main point to be made here is that if the quark thermal distribution ${\cal T}(q_i)$ has the canonical invariant form
\begin{eqnarray}
{\cal T}(q_i) = q_i{dN_q\over dq_i} = Cq_ie^{-q_i/T},     \label{5}
\end{eqnarray}
where $C$ has the dimension of inverse momentum,
then the $\delta$-functions in the RFs require that the hadron \dis s $p^0dN^h/dp_T$ in Eqs.\ (1) and (2) have the common exponential factor, $\exp(-p_T/T)$, for both $h=\pi$ and $p$.  The factors before the  exponentials are different. The integrals in (1) and (2) must yield on dimensional grounds $C^2\pt^2$ and $C^3\pt^3$, respectively, apart from different multiplicative constants. Upon dividing the results of the integration  by  $p_0\pt$ we get the general form
\begin{eqnarray}
{dN_{h}^{\rm TT(T)}\over p_Tdp_T} = {\cal N}_h(\pt)e^{-p_T/T},     \label{6}
\end{eqnarray}
where, for $y=0$, we set $p_0=\pt$ for pion and $p_0=m_T$ for proton, so that
\bq
 {\cal N}_{\pi}=N_0^{\pi} C^2,  \qquad  {\cal N}_p(\pt)=N_0^pC^3{\pt^2\over m_T}, \qquad m_T=(\pt^2+m_p^2)^{1/2},   \label{7}
\eq
$N_0^{\pi}$ and $N_0^p$ being constants.
  Note that the factor $p_T^2/m_T$ in the proton spectrum  causes the $p/\pi$ ratio to vanish as $p_T \rightarrow 0$ on the one hand, but to become large, as $p_T$ increases, on the other.  When $p_T$ exceeds 2 GeV/c, shower partons become dominant  and the above description must be corrected by the effects of thermal-shower  recombination that limits the increase of the $p/\pi$ ratio to a maximum of about 1 \cite{hy}.

  Remaining in the low-$\pt$ region, $\pt<2$ GeV/c, we want to demonstrate that a common value of $T$ is shared by $dN^h/\pt d\pt $ for both $h=\pi$ and $p$. The normalization factor ${\cal N}^h(\pt)$ in Eq.\ (\ref{6}) depends on centrality, which is a subject discussed in the Appendix. Here we consider a specific centrality, 20-30\%, and fit the $\pt$ dependence of the proton spectrum using Eqs.\ (\ref{6}) and (\ref{7}) with free adjustment of the normalization constant. Figure 1 shows the result upon using
\bq
T=0.283\ {\rm GeV}.  \label{8}
\eq
 The one-parameter fit (apart from normalization) is evidently very good compared to the data from Ref.\ \cite{sa}. It demonstrates that the proton is produced in that $\pt$ range by thermal partons and that the flattening of the spectrum at low $\pt$ is due to the prefactor $\pt^2/m_T$ arising from proton recombination.

\begin{figure}[tbph]
\includegraphics[width=.6\textwidth]{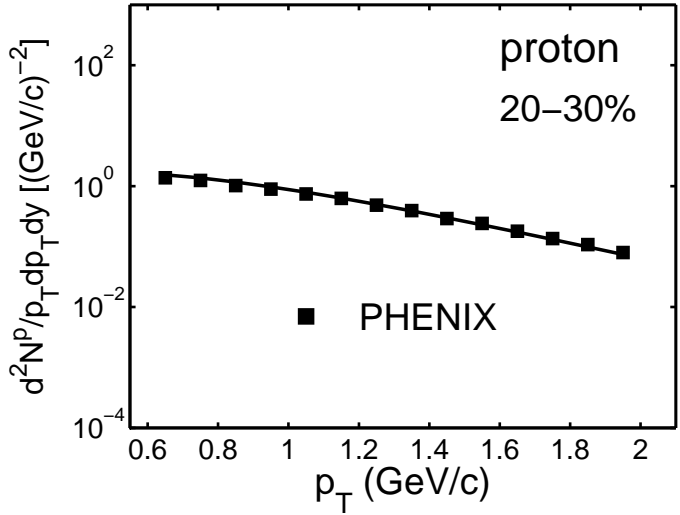}
\caption{Proton spectrum at $y\approx 0$ averaged over $\phi$ (hence, no $1/2\pi$ factor) at 20-30\% centrality. The solid line is a fit of the data by Eqs.\ (\ref{6}) and (\ref{7}) using $T=0.283$ GeV with free adjustment of normalization. The data are from Ref.\ \cite{sa}.}
\end{figure}

As already discussed at the end of the preceding section, the thermal parton distribution we consider is for the time just before hadronization. $T$ in Eq.\ (\ref{5}) is the inverse slope that we have determined here phenomenologically without the assumption that hydro description is appropriate for the entire period from collision to hadronization. It is, however, assumed that local equilibration is achieved for the soft sector at late time to justify the use of Eq.\ (\ref{5}) for all $p_T<2$ GeV/c.
  We refer to $T$ as inverse slope, instead of temperature, because we allow the possibility that the value of $T$ can be affected by the motion of the collective system and become larger than the temperature defined in the local rest frame. For that reason the value of $T$ should not be identified with what is referred to as freeze-out temperature in some fluid description.

Having determined $T$, we have no more freedom to adjust the exponential behavior of  the pion spectrum $dN_{\pi}/\pt d\pt$.  We show in Fig.\ 2 the data from PHENIX \cite{sa} on the pion \dis\ for 20-30\% centrality; the $\exp(-\pt/T)$ factor is shown by the solid line, the normalization being adjusted to fit.
The excellent agreement  thus supports the assertion that both proton and pion spectra can be described by the same $T$ in the exponential factor, $\exp(-\pt/T)$. For $\pt<1$ GeV/c the pion spectrum is dominated by the
resonance contribution which we cannot calculate for lack of knowledge about the RFs of hadrons above the ground states with orbital excitation. For that reason we show only the data in the region $\pt>1$ GeV/c, which is sufficient to verify the commonality of $T$ in the calculable part of our approach.

\begin{figure}[tbph]
\includegraphics[width=.6\textwidth]{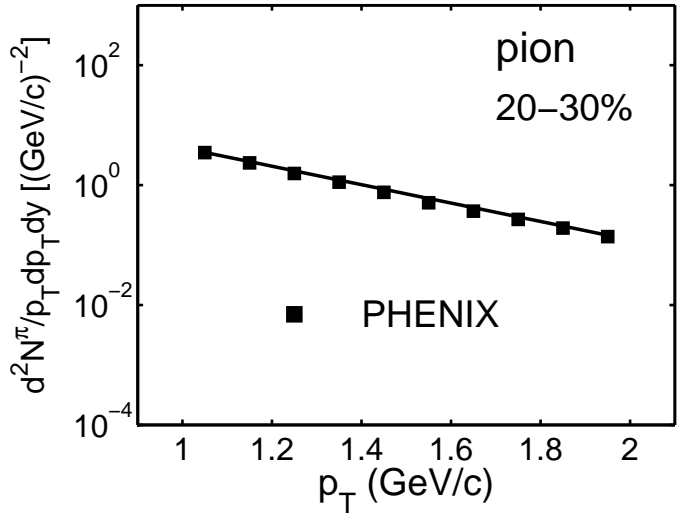}
\caption{The data for the pion spectrum are from Ref.\ \cite{sa}. The solid line that shows $e^{-\pt/T}$ has the same $T$ as for proton. }
\end{figure}

Our basic formulas for recombination shown in Eqs.\ (\ref{1}) and (\ref{2}) are valid for the direct production of all hadrons from thermal partons at any $p_T$. However, those hadrons are not the only ones produced in certain $p_T$ regions due to other processes, such as TS and SS recombination, which dominate at $p_T>2$ GeV/c. For proton production TTT recombination is prevalent at $p_T<2$ GeV/c. For pions from resonance decays that are important for $p_T<1$ GeV/c, Eq.\ (\ref{1}) is inadequate to describe them. When the contributions from shower partons are  significant even at $p_T<2$ GeV/c, appropriate equations will be given below to describe them. $p_T=2$ GeV/c is the upper limit of what we consider in this paper.

\section{Azimuthal Anisotropy without Hydrodynamics}

We now broaden our consideration to include $\phi$ dependence.  For non-central collisions the almond-shaped average initial configuration leads to $\phi$ anisotropy.  The conventional description in terms of hydrodynamics relates the momentum anisotropy to the variation of pressure gradient at early times upon equilibration \cite{kh}.  The success in obtaining the large $v_2$ as observed gives credibility to that approach.  Our alternative approach can be justified  on the same basis that we can also reproduce the empirical $v_2$, as we shall show. Concerns about higher harmonics are at a higher level of details which we shall also address. Our main objective is first and foremost to demonstrate that the essential phenomenological features of hadron production can be reproduced by considering the effects of semihard partons without explicitly treating the fluid flow.

To  include the azimuthal $\phi$ dependence,
let us use $\rho^h(p_T,\phi,b)$ to denote the single-particle distribution of hadron $h$ produced at mid-rapidity in heavy-ion collision at impact parameter $b$,  i.e.,
\begin{eqnarray}
\rho^h(p_T,\phi,b) = {dN_h\over p_Tdp_Td\phi}(N_{\rm part}),     \label{9}
\end{eqnarray}
where $N_{\rm part}$ is the number of participants related to $b$ in the known way through Glauber description of nuclear collision \cite{mrss}.
The main point of our approach is to assert that $\rho^h$  at low $\pt$ can be separated into three components
\bq
\rho^h(p_T,\phi,b) =B^h(p_T,b) + R^h(p_T,\phi,b) + M^h(p_T,\phi,b),     \label{10}
\end{eqnarray}
referred to as Base, Ridge and Minijet components, respectively.
$B^h(p_T,b)$ is azimuthally isotropic, while
 $R^h(p_T,\phi,b)$ and $M^h(p_T,\phi,b)$ are  $\phi$ dependent.
 The first two components are due to the recombination of thermal partons (TT for pion and TTT for proton), while the third is due to thermal-shower recombination (TS and TTS) \cite{hy}. The latter is dominant in the intermediate $\pt$ region ($2<\pt<6$ GeV/c), but is not negligible at low $\pt$ \cite{hz}. In a figure later on in this paper (Fig.\ 6) the relative magnitudes of the three terms will be shown. Because of the smallness of $M^h(p_T,\phi,b)$ relative to the other two for $\pt<2$ GeV/c we shall ignore the shower partons in this and the following sections in order to emphasize the effect of semihard partons on the thermal sector. At the sacrifice of accuracy in reproducing $v_2$, the role of thermal partons in the ridge formation becomes clearer. More accurate result that includes the shower partons will be presented in Sec.\ VI.

Semihard partons created near the surface, and directed outward, can  give rise to $\phi$ anisotropy in the thermal component. That is because each such semihard parton loses
 some energy to the medium, thereby enhancing the thermal motion of the soft partons near its trajectory. Those thermal partons eventually lead to hadrons at late time that are dependent on the azimuthal angle of the semihard parton. In Refs.\ \cite{rch,chy,hz} those hadrons are identified with the ridge that stands above the background with characteristic peaking in $\phi$, which is described by $R^h(\pt,\phi,b)$ in Eq.\ (\ref{10}) with improved treatment to be detailed below. The recoil semihard partons being directed inward are absorbed and randomized.  They become a part of the medium consisting of all  the soft and semihard partons that are farther away from the surface and are unable to lead to hadrons with distinctive $\phi$ dependence.  That is described by the  $B^h(p_T,b)$ component, which should not be confused with the $\phi$-dependent bulk distribution in hydro description. For that reason we avoid using the term bulk.

Ridge is a phenomenon characterized by an extended range in pseudorapidity $\eta$ and a narrow range in $\phi$ \cite{jp}. It may seem hard to relate the ridge to the term $R^h(\pt,\phi,b)$ in Eq.\ (\ref{10}) that has no $\eta$ dependence. It is therefore important to emphasize that we are addressing here the $\phi$ dependence of the ridge at small $\eta$, leaving aside the issues related to the range in $\eta$ that have been considered in our approach in Refs.\ \cite{ch,ch1}. More discussions specifically on the ridge characteristics are given in Sec.\ V below.

Earlier it was found that the azimuthal correlation between a semihard parton and a ridge hadron formed by TT recombination can be described by a Gaussian distribution in $\phi$ with a width $\sigma=0.33$ in order to reproduce the ridge data \cite{ch2}. That result was then extended in a study of the dependence of the ridge yield per trigger on the trigger angle $\phi_s$ relative to the reaction plane \cite{hz,af}. The key piece of physics that succinctly captures the essence of the correlation involving either a trigger or an untriggered semihard parton that generates the ridge is a quantity called $S(\phi,b)$. It is the segment on the initial ellipse through which semihard partons should be emitted if it is to contribute to the formation of any ridge particle that is directed at $\phi$. The importance of $S(\phi,b)$ is that it relates the  spatial and momentum asymmetries.  The derivation of $S(\phi,b)$ given in \cite{hz} is based on the simple geometry of the average initial configuration taken to be an ellipse with width $w$ and height $h$, where $w=1-b/2$ and $h=(1-b^2/4)^{1/2}$ in units of nuclear radius $R_A$. The fluctuations from that configuration will be considered later, but for now it is more important to focus on the relationship between the spatial asymmetry (the short side of the ellipse being on the $x$ axis) and momentum asymmetry of the emitted hadron ($\left<p_x\right> > \left<p_y\right>$).
 Without repeating the derivation  here, let us first state the result, and then follow  up with a discussion on the physics involved.
 The result is
 \begin{eqnarray}
S(\phi,b) = h[E(\theta_2,\alpha) - E(\theta_1,\alpha)],     \label{11}
\end{eqnarray}
where $E(\theta_i,\alpha)$ is the elliptic integral of the second kind with $\alpha=1-w^2/h^2$ and
\begin{eqnarray}
\theta_i = \tan^{-1} \left({h\over w}\tan\phi_i \right), \quad \phi_1 = \phi - \sigma, \quad \phi_2 = \phi + \sigma,     \label{12}
\end{eqnarray}
for $\phi_i \leq \pi/2$, and an analytic continuation of it for $\phi_2 > \pi/2$.  Thus $S(\phi,b)$ is a compact formula that is completely calculable for any given $b$, and has the property that it decreases as $\phi$ is increased from 0 to $\pi/2$, more so at medium or large $b$ than at small $b$. Since it will prescribe the $\phi$ dependence of
 $R^h(p_T,\phi,b)$, the average hadron momentum for non-central collisions is larger along $p_x$ than along $p_y$.

It should be noted that Eqs.\ (\ref{11}) and (\ref{12}) involve azimuthal angles only, which are meaningful in both the coordinate space and momentum space. The relationship between the angles is based on the correlation between the hadronic momentum in the ridge at angle $\phi$ and the direction $\phi_s$ of the semihard parton emitted from the ellipse in the coordinate space with $|\phi-\phi_s|<\sigma$. Although the angles $\phi_s$ and $\phi$ are at early and late times, respectively, they are nevertheless correlated, since $\sigma$ was determined in Ref.\ \cite{ch2} to fit the data on ridge yield as a function of the trigger angle $\phi_s$ \cite{af}. The only assumption here is that the trigger angle is identified with the semihard parton angle that is not directly measurable. Such an identification does not rely on the details of hydrodynamics.

The physical origin of the $\phi$ anisotropy generated by semihard partons is that on the broader side of the spatial ellipse  there can be more semihard partons within an angle $\sigma$ contributing to a hadron emitted with small $\phi$ at small $y$,
 where the curvature on the ellipse is small thus allowing a longer segment on the ellipse with normal in the range $\phi\pm\sigma$.
On the narrow side at the tip of the ellipse the curvature is larger thus restricting the segment through which  semihard partons  can contribute to a hadron at $\phi\sim \pi/2$. The mechanism that gives rise to this orthogonality between the spatial and momentum asymmetry axes is entirely different from that in the fluid description which is basically that the higher pressure gradient along the $x$ axis in the initial state generates more hadronic momentum along that direction in the final state. Without hydrodynamics we, of course, cannot describe the evolutionary history of the system.
While the hydro approach assumes rapid thermalization, we allow unspecified time interval for expansion and equilibration except that by the time of hadronization at late time the soft partons have exponential $\pt$ behavior, which is the only property we ascribe to the thermal partons, apart from the $\phi$ dependence of the enhanced thermal partons caused by  the semihard partons on their way out of the medium at early time. If this approach can lead to sensible phenomenology of the azimuthal harmonics, as we shall show below, then it is an alternative that should be weighed against the merits of the conventional approach.

The discussion above is about the $\phi$ anisotropy of the thermal partons. Also to be considered is the role of the shower partons which are the fragmentation products of the semihard parton outside the medium before they hadronize. Inasmuch as the former reveals the effect of the semihard partons on the medium through which they traverse,  the latter is the minijet manifestation of the semihard partons themselves by TS recombination. The hadrons formed are to be described by the third term $M^h(\pt,\phi,b)$ in Eq.\ (\ref{10}). The SS component is not considered in this paper because it is negligible at $\pt<2$ GeV/c \cite{hy,hy2}. Shower partons can arise from semihard and hard partons created throughout the medium in random directions. Because of jet quenching the partons that emerge from the medium have reduced momenta, and the distribution of the shower partons generated by subsequent fragmentation peaks at low $\pt$. They recombine with the thermal partons in the immediate vicinity of the emerging partons and therefore form hadrons that have approximately the same $\phi$ angles as the initiating semihard or hard partons. Upon averaging over all events the azimuthal dependence of the TS term can have all harmonic components as in Fourier decomposition. Since the $\pt$ dependence of the azimuthal harmonics is what we shall calculate and compare with data, we summarize here the formulas for pion production by TS recombination that are relevant
\bq
{dN_{\pi}^{\rm TS}\over p_Tdp_T} = {2\over p_T^2} \int {dp_1\over p_1}{dp_2\over p_2} {\cal T}(p_1){\cal S}(p_2,\bar \xi){\cal R}_{\pi}(p_1,p_2,p_T),     \label{13}
\end{eqnarray}
where
\begin{eqnarray}
{\cal S}(p_2,\bar \xi)=\int {dq\over q}\sum_i \bar F_i(q,\bar \xi) S_i(p_2/q).  \label{14}
\end{eqnarray}
${\cal S}(p_2,\bar \xi)$ is the shower parton \dis\ integrated over all semihard parton $q$ at the medium surface after momentum degradation parametrized by $\bar\xi$, and $S_i(p_2/q)$ is the \dis\ of shower partons with momentum $p_2$ in a jet of type $i$ with momentum $q$. The details of these quantities can be found in Refs.\ \cite{hy,hy2}.

\section{Second Harmonic of $\phi$ Anisotropy}
This topic is usually referred to as elliptic flow, a terminology that is rooted in hydrodynamics.  Since hydro is not the basis of our investigation, we use the more general language of harmonic analysis and refer to $v_n$ as the harmonic coefficients
\begin{eqnarray}
v_n^h(p_T,b) = \langle \cos n\phi \rangle_{\rho}^h = {\int_0^{2\pi} d\phi \cos n\phi\rho^h(p_T,\phi,b)\over \int_0^{2\pi} d\phi\rho^h(p_T,\phi,b)},     \label{15}
\end{eqnarray}
where $\rho^h(\ppb)$ in our formalism has the three components given in Eq.\ (\ref{10}). We now describe the $\phi$ dependence of $R^h(\ppb)$ and $M^h(\ppb)$ separately.

As discussed in the last section, $R^h(\ppb)$ contains the $\phi$ anisotropy arising from the initial elliptical spatial configuration through the $S(\pb)$ function that transforms the spatial to momentum asymmetry. We now insert some details omitted in our general discussion. Since the elliptical axes need not coincide with the reaction plane that contains the impact parameter vector $\vec b$, we introduce a tilt angle $\psi_2$ and average over it. Furthermore, we modify the notation slightly by using $S_2(\pb)$ to denote what is defined in Eq.\ (\ref{11}) and write the average over $\pi/2n$ as
\bq
\tilde S_2(\pb)={2\over \pi}\int_{-\pi/4}^{\pi/4} d\psi_2 S_2(\phi-\psi_2, b) .  \label{16}
\eq
We then define $S(\pb)$ as the normalized $\tilde S_2(\pb)$, i.e.,
\bq
S(\pb)=\tilde S_2(\pb)\left /{{1\over 2\pi} \int_0^{2\pi} d\phi \tilde S_2(\pb)}\right. . \label{17}
\eq
 Following our discussion in the last section on the ridge component of $\rho^h$ that responds to the minijets through TT recombination, we now can write
\bq
R^h(\ppb)=S(\pb) \bar R^h(\pt,b) ,  \label{18}
\eq
where $\bar R^h(\pt,b)$ is the second of two components of $dN_h^{\rm TT(T)}/\pt d\pt$. The exponential behavior of the first component, which is the $\phi$-independent base component $B^h(\pt,b)$, has  a lower $T_0$ than the overall $T$  for the sum of the two thermal terms described by Eq.\ (\ref{6}). Thus, with the unenhanced base thermal component expressed as
\bq
B^h(\pt,b) ={\cal N}_h(\pt,b)e^{-\pt/T_0},  \label{19}
\eq
the enhanced ridge component is
\bq
\bar R^h(\pt,b)= {\cal N}_h(\pt,b)[e^{-\pt/T} -e^{-\pt/T_0}].  \label{20}
\eq
We emphasize that the only factor that  depends on the hadron type is $ {\cal N}_h(\pt,b)$. It is a specific property of the recombination model that the exponential factors of the hadrons (whether $\pi$ or $p$) are inherited from those of the partons as discussed in the  preceding section. Note also that $T_0$ is the only unknown parameter introduced here. If for the present we neglect the TS component for the sake of simplicity, since it is small at low $\pt$, we would have only the first two terms of $\rho^h(\ppb)$ in Eq.\ (\ref{10}), and the formalism up to this point should be sufficient to provide an approximate description of the second harmonic.

Applying Eq.\ (\ref{18}) to (\ref{15}), we obtain for $n=2$
\begin{eqnarray}
v_2^h(p_T,b) &=& {\bar R^h(\pt,b){1\over 2\pi}\int_0^{2\pi} d\phi \cos2\phi S(\phi,b) \over B^h(\pt,b)+\bar R^h(\pt,b)}   \nonumber \\
&=& {\langle \cos2\phi \rangle_S\over Z^{-1}(p_T) + 1},     \label{21}
\end{eqnarray}
where
\begin{eqnarray}
\langle \cos2\phi \rangle_S &=& {1\over 2\pi} \int_0^{2\pi} d\phi \cos2\phi S(\phi,b),     \label{22} \\
Z(p_T) &=& {\bar R^h(\pt)\over B^h(\pt)}= e^{p_T/T'} - 1, \qquad T'={T_0 T\over T- T_0} .     \label{23}
\end{eqnarray}
These equations are remarkable in that the $b$ dependence resides entirely in Eq.\ (\ref{22}) and the $p_T$ dependence entirely in Eq.\ (\ref{23}); furthermore, there is no explicit dependence on the hadron type.

\begin{figure}[tbph]
\centering
\hspace*{-8.8cm}
\includegraphics[width=0.47\textwidth]{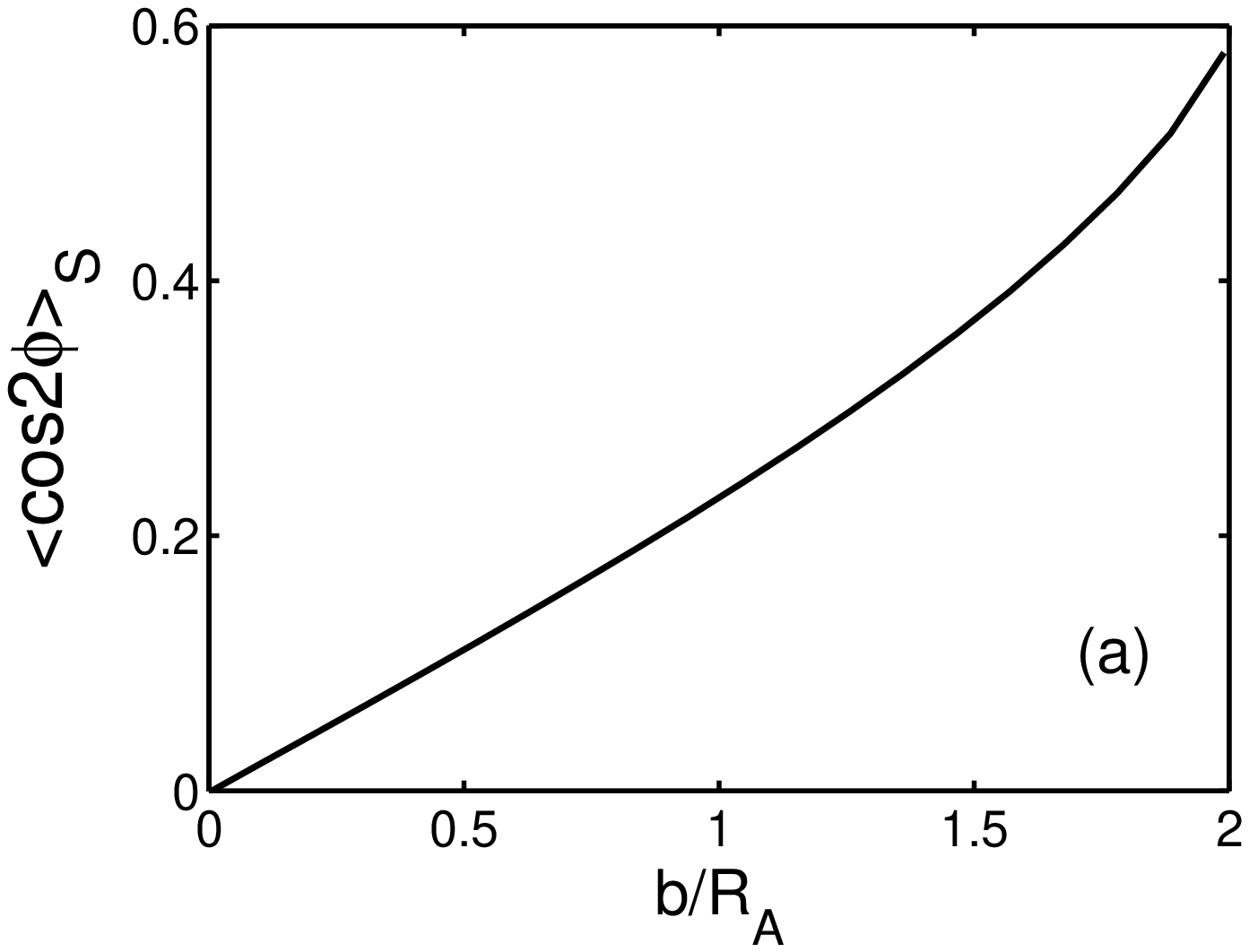}

\centering
\vspace*{-6.cm}
\hspace*{8.cm}
\includegraphics[width=0.53\textwidth]{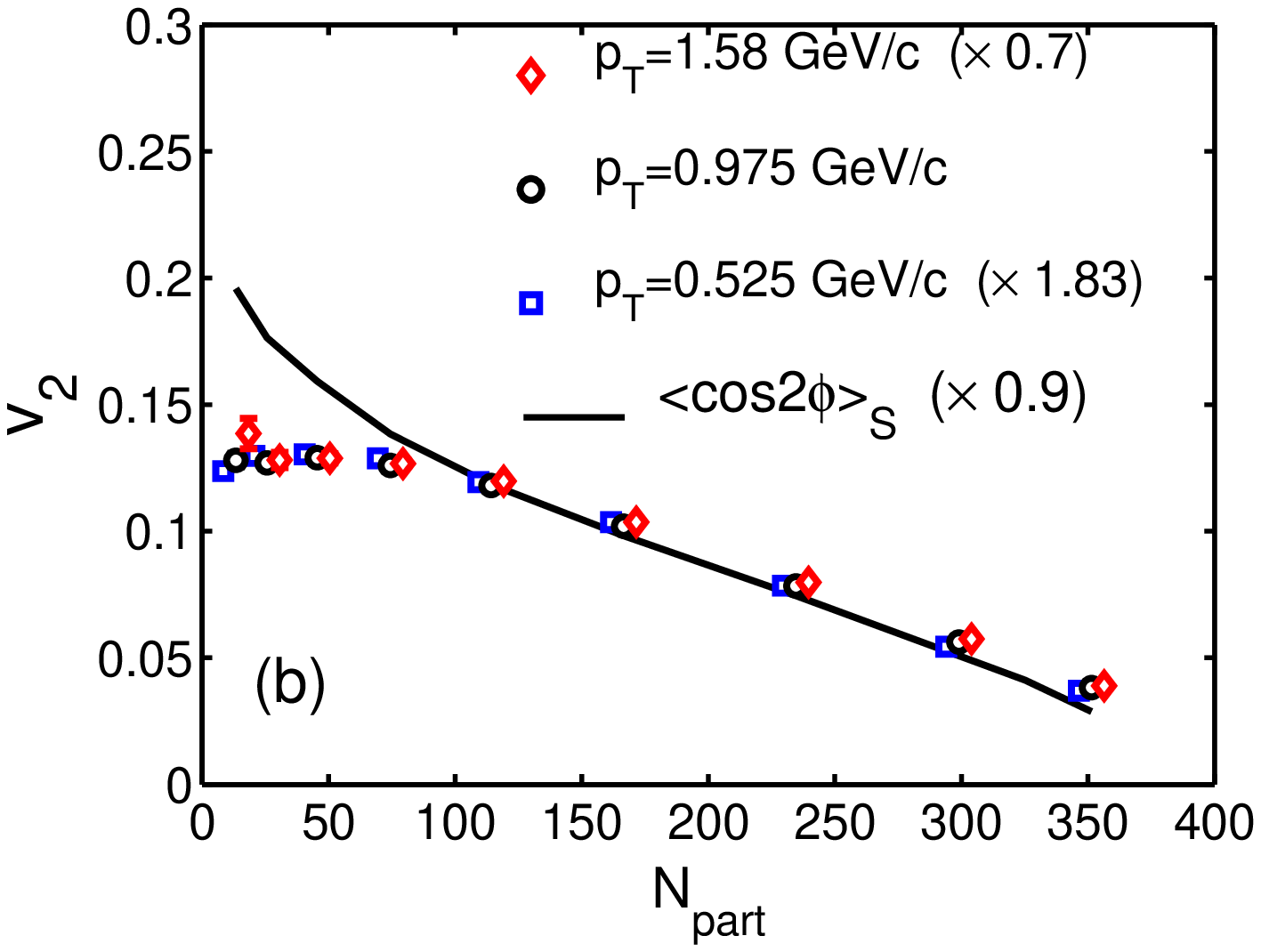}

\caption{(Color online) (a) Average of $\cos2\phi$ weighted by $S(\phi,b)$ vs  impact parameter $b$ in units of $R_A$. (b) Common dependence of $v_2^h(\pt,b)$ on $N_{\rm part}$ for various $\pt$, shifted vertically for comparison. The diamond and square points are horizontally shifted slightly from the points in circles to aid visualization. The solid line is from $\left<\cos2\phi\right>_S$ shown in (a), but rescaled and plotted in terms of $N_{\rm part}$. The data are from Ref.\ \cite{ja}.}
\end{figure}

From Eqs.\ (\ref{11}), (\ref{16}), (\ref{17}) and (\ref{22}) we can calculate $\left< \cos 2\phi\right>_S$ and obtain its dependence on $b$.  The result is shown in Fig.\ 3(a). To check how realistic  phenomenologically the factorizability of $\pt$ and $b$ dependencies of $v_2^h(\pt,b)$ is, we show  in Fig.\ 3(b) the data from Ref.\ \cite{ja} on $v_2^h(\pt,N_{\rm part})$ for three $\pt$ values, but shifted vertically so that they agree with the data for $\pt=0.975$ GeV/c for most of large $N_{\rm part}$. The diamond and square points are slightly shifted horizontally to spread out the overlapping points for the sake of visual distinguishability. The fact that their dependencies on $N_{\rm part}$ are so nearly identical is remarkable in itself. The solid line is a reproduction of the curve in Fig.\ 3(a) but plotted in terms of $N_{\rm part}$, and reduced in normalization by a factor 0.9 to facilitate the comparison with the data points. For $N_{\rm part}>100$ the line agrees with the data on $v_2$ very well, thus proving the factorizability  of Eq.\ (\ref{21}). For $N_{\rm part}<100$, corresponding to $b/R_A>1.3$ or centrality $>40$\%, there is disagreement which is expected because the density is too low in peripheral collisions to justify the simple formula in Eq.\ (\ref{11}) that is based on no punch-through of recoil partons.

Since $T_0$ describes the $\pt$ dependence of the   $B^h(\pt,b)$ component,
it  is not directly related to any observable spectrum.  Thus we turn to $v_2^h(\pt,b)$ in Eq.\ (\ref{21}) for pion first and find that the low-$\pt$ data of $v_2^{\pi}(\pt,b)$ can be well reproduced. In Fig.\ 4(a) is shown the data for pion from Ref.\ \cite{ja} for 0-5\% centrality; the solid line is the result of our calculation from Eq.\ (\ref{21}) using
\bq
T_0=0.245\ {\rm GeV} .  \label{24}
\eq
The fit though not perfect is remarkable because the normalization of $v_2^\pi$ is fixed by Eq.\ (\ref{21}) without freedom of adjustment. Note that we have not used any more parameters besides $T_0$ to accomplish this, which is a fitting procedure not more elaborate than the hydro approach where the initial condition and viscosity are adjusted. A better result will be shown below when the TS component is taken into account.

\begin{figure}[tbph]
\centering
\hspace*{-8.4cm}
\includegraphics[width=0.5\textwidth]{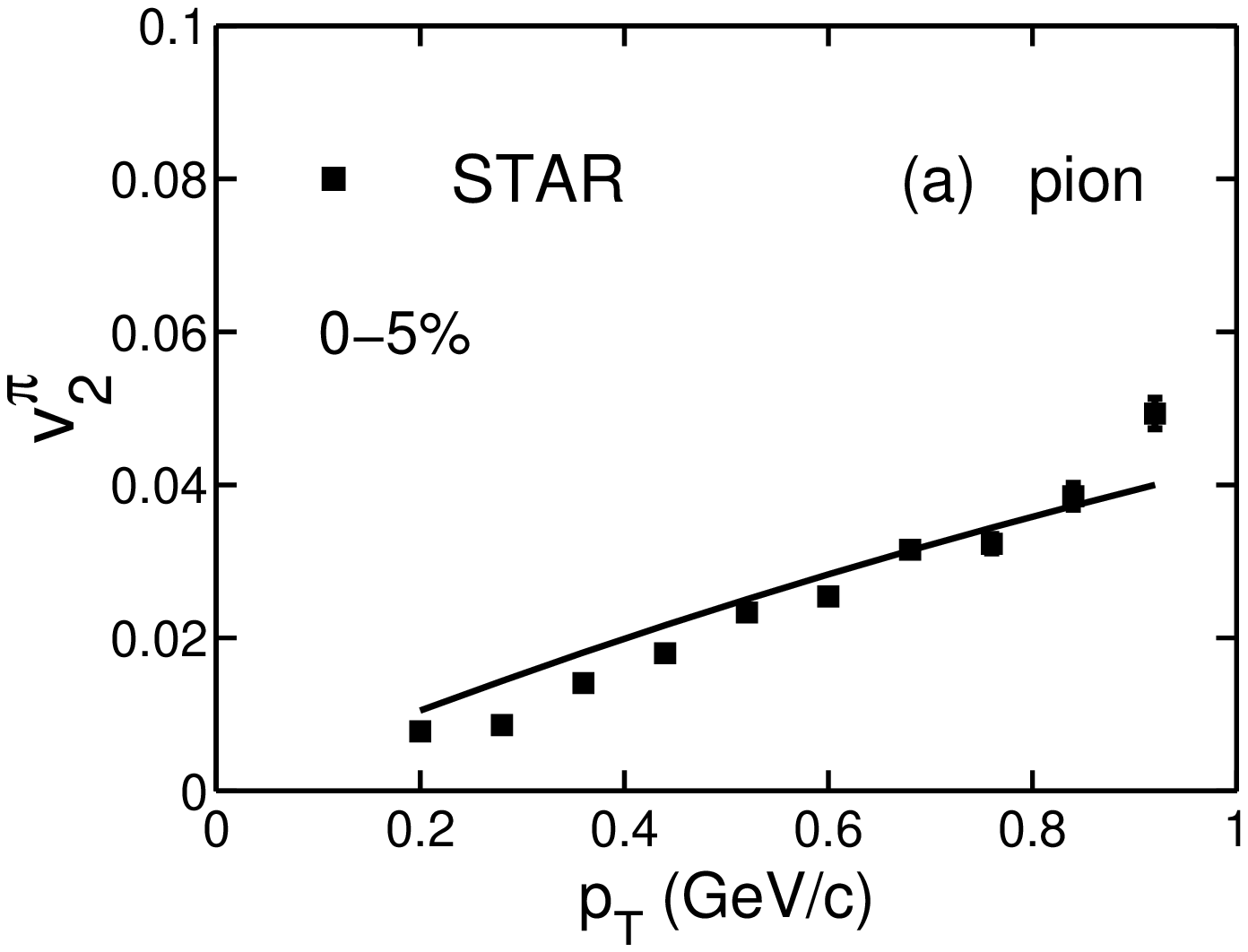}

\centering
\vspace*{-6.2cm}
\hspace*{8.2cm}
\includegraphics[width=0.5\textwidth]{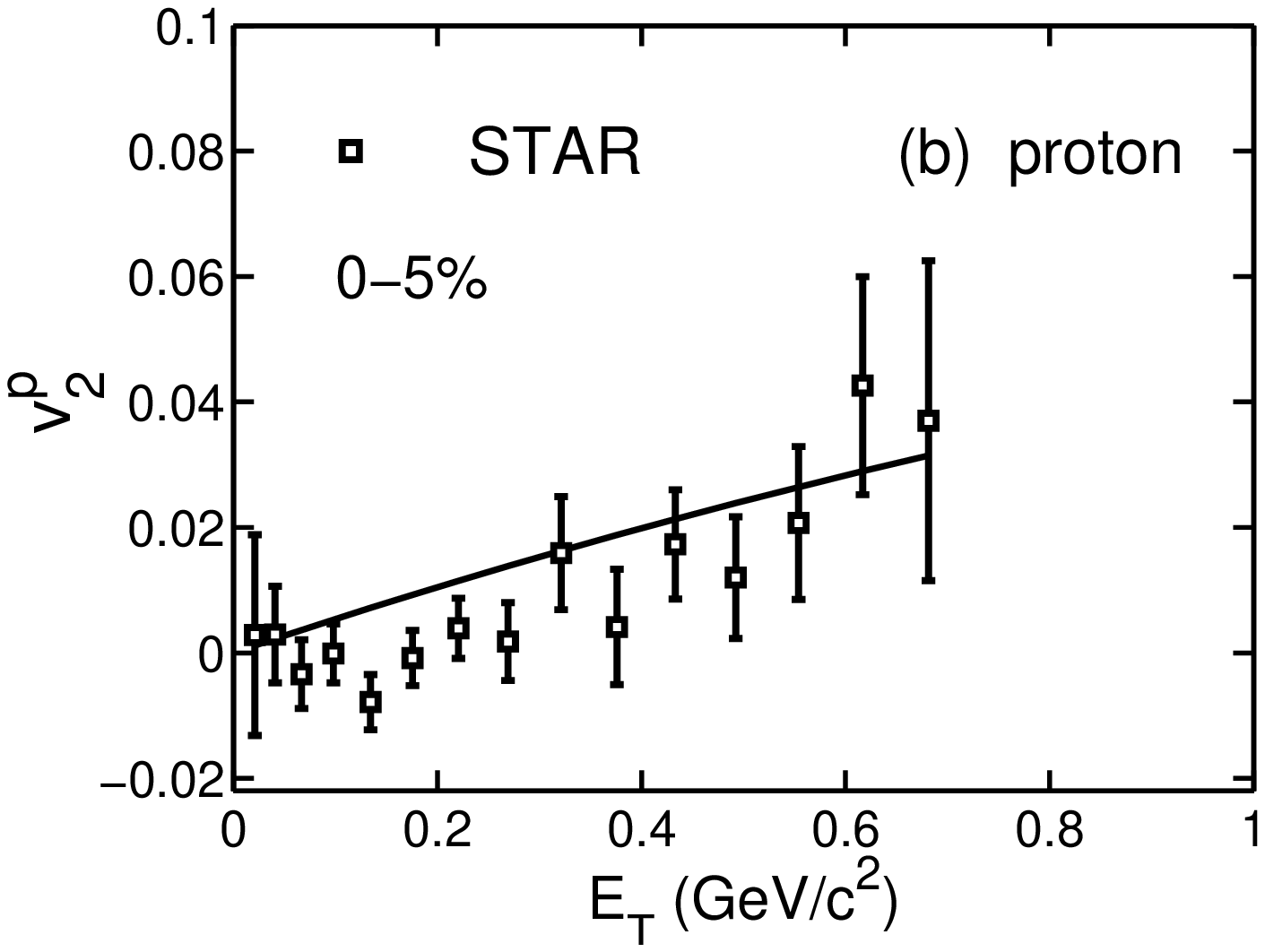}
\caption{$v_2^\pi$ at 0-5\% centrality for (a) pion and (b) proton. The data are from Ref.\ \cite{ja}. The solid lines are calculated from Eq.\ (\ref{21}) using $T_0$ as the adjustable parameter to fit $v_2^\pi$.}
\end{figure}

For proton $v_2^p(\pt,b)$ we take the mass effect into account in the same way as we have done before \cite{chy} by working with transverse kinetic energy $E_T$, where
\bq
E_T(p_T)=m_T(\pt)-m_h,   \label {25}
\eq
and adopt the ansatz that $\pt$ is to be replaced by $E_T$ in Eq.\ (\ref{23}), i.e.,
\bq
Z(\pt)=e^{E_T(\pt)/T'}-1 .  \label{26}
\eq
Using the same formula Eq.\ (\ref{21}) without any change of $T_0$,  we show in Fig.\ 4(b) that the result agrees with the proton data \cite{ja} fairly well.

So far we have concentrated on $v_2^h(\pt,b)$ at low $\pt$ for 0-5\% centrality as a first step toward demonstrating the phenomenological relevance of our approach based on thermal partons only in the first two terms of $\rho^h(\ppb)$ in Eq.\ (\ref{10}). To widen the $\pt$ and $b$ ranges, as well as to consider higher harmonics, we must include the third component generated by TS recombination. But before doing so, we pause in our study of the azimuthal anisotropy and revisit the subject of ridge.

\section{The Ridge}

As remarked earlier in Sec.\ III, our study of the $\pt$ and $\phi$ dependencies of $\rho^h(\pt,\phi,b)$ at midrapidity in this paper does not give us the scope that can include the rapidity dependence. The $\Delta\eta$ range of the ridge either in triggered correlation or in untriggered autocorrelation reaches as high as 4 \cite{ja1,bia,ba2,bia1,md} and has therefore been regarded as long-range correlation \cite{ad,gmm,dgl}. That problem in the framework of our approach is addressed in Ref.\ \cite{ch1}. The subject of our concern here is the property of the ridge at $\eta\sim 0$; more specifically, we describe the effect of ridge in the inclusive \dis\ at low $\pt$.

Ignoring the third term in Eq.\ (\ref{10}) for the present discussion, and using Eq.\ (\ref{18}) for the second term, we have
\bq
\rho^h(\ppb)=B^h(\pt,b)+S(\pb) \bar R^h(\pt,b)     \label{5.1}
\eq
so that upon averaging over $\phi$, we obtain the two terms
\bq
\bar\rho^h(\pt,b)=B^h(\pt,b)+\bar R^h(\pt,b) .    \label{5.2}
\eq
Their sum is the inclusive \dis\ with the exponential behavior given by Eq.\ (\ref{6}) for $\pt<2$ GeV/c, whereas they separately behave according to Eqs.\ (\ref{19}) and (\ref{20}). $B^h(\pt,b)$ has been referred to as base, while $\bar R^h(\pt,b)$ describes the ridge. They are both the hadronic products of the recombination of thermal partons.

It is not obvious by examining Eq.\ (\ref{20}) that $\bar R^h(\pt,b)$ exhibits ridge structure, but the derivation of $S(\pb)$ outline in Sec.\ III clearly indicates that $R^h(\ppb)$ has the quadrupole behavior because of the effect of semihard partons. That is, in addition to the $\phi$-independent base term, the additional ridge term is made manifest at $\phi$ when semihard partons are within a cone of width $\sigma$ around $\phi$, owing to the enhancement of the thermal partons in the cone due to energy loss by the semihard partons. The hadrons formed in the ridge has a higher $\left<\pt\right>$ than those in the base. For a single-particle \dis\ we have, of course, no trigger to select a direction around which the enhancement can be measured. But that does not mean that the effect of semihard partons (and therefore the ridge) is not present in the inclusive \dis. For $b/R_A=1$ we show in Fig.\ 5 the $\pt$ dependencies of  $B^\pi(\pt,b)/{\cal N}_\pi(\pt,b)$  and $\bar R^\pi(\pt,b)/{\cal N}_\pi(\pt,b)$, defined in Eqs.\ (\ref{19}) and (\ref{20}),  by the (red) dash-dotted line and (blue) dashed line, respectively, and referred to as base and ridge. The former has $T_0=0.245$ GeV according to Eq.\ (\ref{24}); the latter is not exactly straight in Fig.\ 5 but can be fitted by $\exp(-p_T/T_R)$ with
\bq
T_R=0.32\ {\rm GeV}.   \label{5.3}
\eq
This larger inverse slope clearly indicates that the hadrons in the ridge are the products of enhanced thermal partons compared to those in the base. The sum $\bar\rho^\pi(p_T,b)/{\cal N}_\pi(\pt,b)$,  which is the inclusive, is shown by the (black) solid line in Fig.\ 5, whose inverse slope is given by $T$ in Eq.\ (\ref{8}). Note that $T_R-T=47$ MeV is very close to the value 45 MeV that Putschke reported as the difference in the values of  $T$ between the triggered ridge and inclusive \dis s \cite{jp}.

\begin{figure}[tbph]
\includegraphics[width=.6\textwidth]{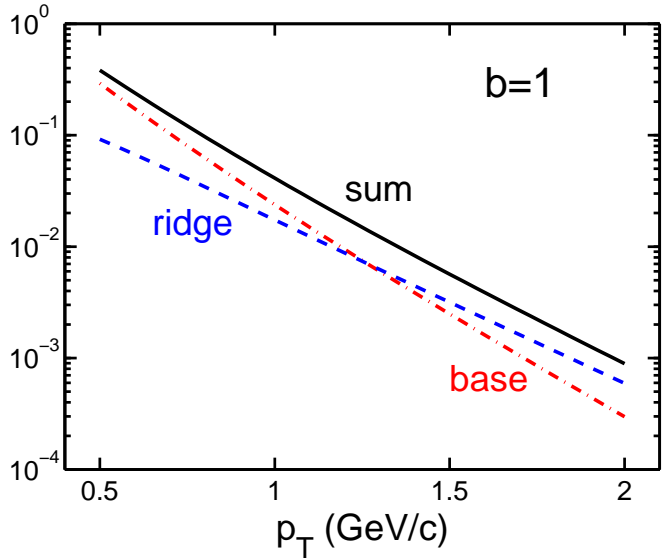}
\caption{(Color online) The $\pt$ \dis s of the base $B^\pi(\pt,b)/{\cal N}_\pi(\pt,b)$  in red dash-dotted line, ridge $\bar R^\pi(\pt,b)/{\cal N}_\pi(\pt,b)$ in blue dashed line and their sum in black solid line. The centrality is for $b=1$ in units of $R_A$. }
\end{figure}

Figure 5 does not show the $\eta$ and $\phi$ characteristics of the ridge, but the $\phi$ dependence of $R^\pi(\ppb)$ is completely contained in $S(\pb)$ as expressed in Eq.\ (\ref{18}). Hence, $v_2(\pt,b)$ and the ridge are intimately related, both being the consequences of semihard partons. If the shower partons generated by the semihard partons lead to a trigger particle, then the hadrons associated with that trigger would exhibit a peak in $\phi$ around the trigger direction, as was shown in Ref.\ \cite{ch2}, in agreement with data \cite{af}. In fact, a prediction on the asymmetry of hadrons produced on the two sides of the trigger direction was subsequently verified to exist in the data \cite{jrk}. If the trigger direction is integrated over, then the $\phi$ \dis\ of the ridge hadrons would behave as $R^\pi(\pt,\phi,b)$.

\section{Higher Harmonics}

In recent years the study of $\phi$ anisotropy has advanced from $v_2(\pt,b)$ to higher harmonics \cite{aglo,qp,hm,ka}. It is widely accepted that the coefficients $v_n$ with $n>2$ are due to fluctuations of the initial configuration whose spatial eccentricity $\varepsilon_n$ leads to the corresponding $v_n$ of the hadronic momentum \dis\ through hydrodynamical flow \cite{ty,sjg,gglo}. It has therefore become our burden of proof that our approach can also reproduce the higher $v_n$ without flow.

In the preceding section we have shown that $v_2$ can be understood in terms of the $\phi$ dependence of the TT recombination of the thermal partons affected by the passage of semihard partons through the medium. Although we have no transport model to follow the evolution of the system, the space-momentum transformation is accomplished by studying the minijets emitted from the initial elliptical configuration, hence $v_2$. It is then natural for us to focus on the effects of the same minijets on the higher harmonics. In a sense minijets play a role similar to the fluctuations of the initial configuration, because their effects on the $\phi$ \dis\ present a departure from the consequence of the simple and smooth approximation of that configuration by ellipse, except that minijets are themselves fluctuations in the momentum space and do not depend on flow dynamics. In our formalism the minijets affect the low-$\pt$ region through TS recombination.

Since minijets are produced in any given event in unpredictable directions, the average $\phi$ \dis\ can have all terms in a harmonic analysis. The only aspect of the behavior that our formalism has a predictable power is the dependence on $\pt$ and centrality  because the $\phi$-integrated TS component of recombination has already been formulated and parametrized. To be explicit, let us write the third component of $\rho^h(\ppb)$ in Eq.\ (\ref{10}) as
\bq
M^h(\ppb)=J(\pb) \bar M^h(\pt,b), \label{27}
\eq
where $J(\pb)$ describes the $\phi$-dependent part of the minijet contribution, which is assumed to be factorizable from the average $\bar M^h(\pt,b)$ in the same manner as for $R^h(\ppb)$ in Eq.\ (\ref{18}). Moreover, as in Eq.\ (\ref{17}), $J(\pb)$ is the normalized form of $\tilde J(\pb)$
\bq
J(\pb)=\tilde J(\pb)\left/ {{1\over 2\pi} \int_0^{2\pi}d\phi \tilde J(\pb)}\right.,  \label{28}
\eq
where $\tilde J(\pb)$ contains all the harmonic components, $\cos n\phi$, averaged over the tilt angle $\psi_n$, i.e.,
\bq
\tilde J(\pb)=1+b \sum_{n=2}^\infty a_n {n\over \pi}\int_{-\pi/2n}^{\pi/2n} d\psi_n \cos n(\phi-\psi_n).  \label{29}
\eq
The $b$ dependence in the above will be discussed below. We include $n=2$ term in Eq.\ (\ref{29}) as an additional contribution to $v_2$ besides the one from Eq.\ (\ref{18}), which is dominant at low $p_T$ because it is from TT recombination. With the $M^h$ term arising from TS recombination included, we shall go above the $p_T<1$ GeV region shown in Fig.\ 4. There is no way to calculate the amplitudes $a_n, n\ge 2$, but the $\pt$ and $b$ dependence of $\bar M^h(\pt,b)$ is a unique attribute of our model, and will be put to test in our phenomenology of $v_n(\pt,b)$ below.

Including all three components of $\rho^h(\ppb)$ in Eq.\ (\ref{10}), we obtain from (\ref{15})
\bq
v_n^h(\pt,b)={\left<\cos n\phi\right>_S \bar R^h(\pt,b) + \left< \cos n\phi\right>_J \bar M^h(\pt,b) \over \bar\rho^h(\pt,b)}  ,  \label{30}
\eq
where
\bq
\bar\rho^h(\pt,b)&=&B^h(\pt,b)+\bar R^h(\pt,b)+\bar M^h(\pt,b),  \label{31} \\
\left< \cos n\phi\right>_J&=&{1\over 2\pi} \int_0^{2\pi} d\phi \cos n\phi J(\pb).  \label{32}
\eq
$\left<\cos n\phi\right>_S$ is as defined in Eq.\ (\ref{22}) for any $n$, but it is zero for $n\ge 3$ because of the periodicity of $S(\pb)$. Indeed, $\left< \cos n\phi\right>_J$ receives contribution only from the $a_n$ term in Eq.\ (\ref{29}) because of the orthogonality of the harmonics. It is clear
from Eq.\ (\ref{30}) that the $\pt$ and $b$ dependencies of $v_n^h(\pt,b)$ are no longer separable as in Eq.\ (\ref{21}), when $\bar M^h(\pt,b)$ is included as is necessary for $p_T>1$ GeV.

	In the following we shall consider only pion production by TS recombination, since TTS and TSS recombination for proton is less important for $\pt<2$ GeV/c. The equation for $dN_{\pi}^{\rm TS}/\pt d\pt$ given in (\ref{13}) and (\ref{14}) are more elaborate than we need for $\bar M^{\pi}(\pt,b)$. The $b$ dependence of $\bar\xi$ given in \cite{hy2} is of a scaling form at intermediate $\pt$ and is complicated. We shall use the approximate form used in the earlier treatment \cite{hy} where for the most central collisions
\bq
\left.{dN_{\pi}^{\rm TS}\over \pt d\pt}\right|_{b=0}&=&{2C\over \pt^3}\int_0^{\pt}dp_1 p_1 e^{-p_1/T}{\cal S}(\pt-p_1) ,  \label{33}  \\
{\cal S}(p_2)&=&\xi_{\rm eff}\,\sigma_g\int_{k_0}^\infty dk k f_g(k) S_g(p_2/k) ,  \label{34}
\eq
where only gluon jets are considered explicitly with $f_g(k)$ being the \dis\ of (semi)hard gluon created with momentum $k$. The factor $\sigma_g=1.2$ is used to take into account the other (semi)hard partons whose contribution we approximate by adding 20\% to the contribution from gluon jets \cite{hz2}. The parameter $\xi_{\rm eff}$ is the effective fraction of (semi)hard partons created anywhere in the medium that emerges to fragment into clusters of shower partons; it is determined phenomenologically to be 0.07 \cite{hy}. For non-central collisions we regard $\bar M^{\pi}(\pt,b)$ to be proportional to $C(N_{\rm part})N_{\rm coll}(b)$, where $C(N_{\rm part})\propto N_{\rm part}^{0.52}$ [see Eqs.\ (A1) and (\ref{5})]; it is the normalization of the thermal parton \dis. $N_{\rm coll}(b)$ is the number of binary collisions that normalizes $f_g(k)$. We thus  have
\bq
\bar M^{\pi}(\pt,b)\left.={C(N_{\rm part})N_{\rm coll}(b)\over C(N_{\rm part}^{max})N_{\rm coll}(b=0)}{dN_{\pi}^{\rm TS}\over \pt d\pt}\right|_{b=0} , \label{35}
\eq
which completely specifies the $\pt$ dependence. There is just one more piece of physics that needs to be added. That is the decrease of average path length in the medium as the collision becomes more peripheral. Its consequence is that more fraction of the (semi)hard partons can emerge from the medium as $b$ increases. In \cite{hy2} a detailed study of dependence of the nuclear modification factor on $\phi$ and $b$ has been carried out. For the purpose of promoting  a transparent connection between the harmonic coefficients $v_n$ and the input, we make the simple first-order approximation here that the increase of minijets with $b$ can be expressed as a linear rise shown in Eq.\ (\ref{29}), which then exhibits very plainly the $\phi$ and $b$ dependencies of $\tilde J(\pb)$, and therefore also $ M^{\pi}(\ppb)$.

\begin{figure}[tbph]
\includegraphics[width=.6\textwidth]{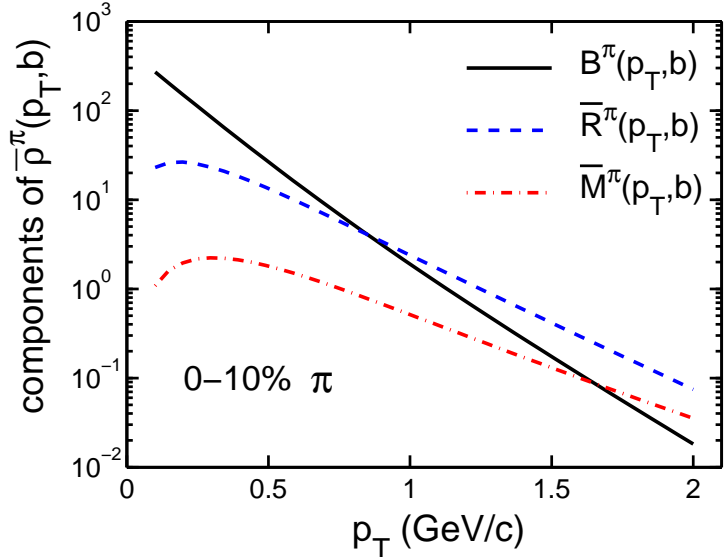}
\caption{(Color online) $ B^{\pi}(\pt,b)$, $\bar R^{\pi}(\pt,b)$ and $\bar M^{\pi}(\pt,b)$ for 0-10$\%$ centrality.}
\end{figure}

It is useful to have a visual comparison of the various components of $\bar\rho^\pi(\pt,b)$ in Eq.\ (\ref{31}). In Fig.\ 6 we show $ B^{\pi}(\pt,b)$, $\bar R^{\pi}(\pt,b)$ and $\bar M^{\pi}(\pt,b)$ for 0-$10\%$ centrality, determined from using Eqs.\ (\ref{19}), (\ref{20}) and (\ref{35}), respectively. Evidently, $\bar R^{\pi}(\pt,b)$ and $\bar M^{\pi}(\pt,b)$ become increasingly more important at increasing $\pt$. They set the scale of $v_n^\pi(\pt,b)$ through their roles in Eq.\ (\ref{30}). For specific harmonics, we limit ourselves to $n=2, 3$ and 4 and calculate $v_2^\pi(p_T,b), v_3^\pi(p_T,b)$ and $v_4^\pi(p_T,b)$. The results are shown by the solid lines in Fig.\ 7, where the data are from PHENIX \cite{aa2}.
 The values of the parameters used are
\bq
a_2=0.6, \qquad a_3=1.6, \qquad a_4=1.2.   \label{36}
\eq
It is remarkable how well the calculated curves agree with the data in $p_T$ dependence for four centrality bins in each case.
 One parameter $a_n$ for each $n$ can affect only the magnitude of $v_n(\pt,b)$, so the excellent reproduction of the $\pt$ and $b$ dependencies reveals the basic attributes of the approach that we have taken to describe the harmonics.

\begin{figure}[tbph]
\includegraphics[width=0.55\textwidth]{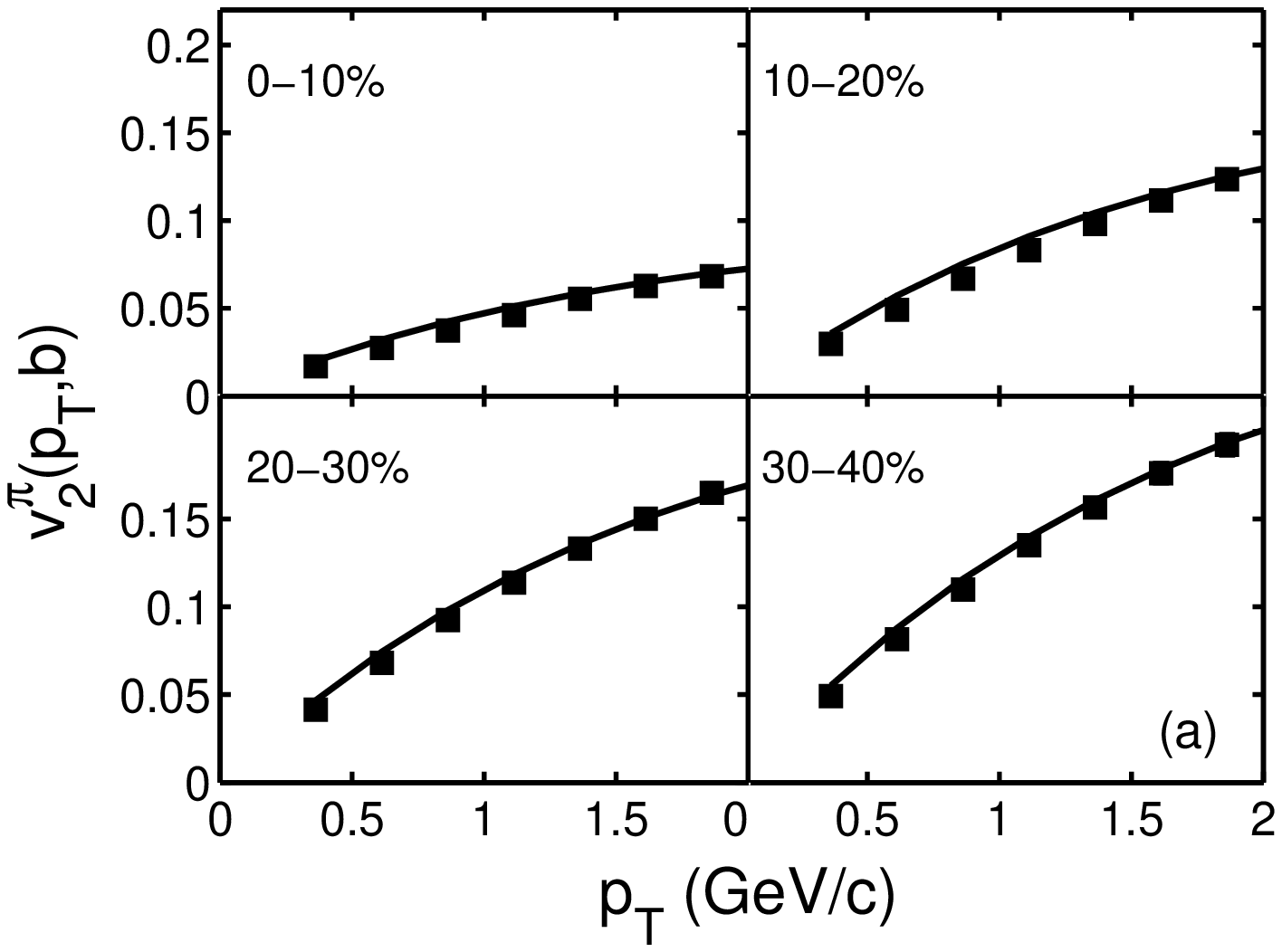}

\includegraphics[width=0.55\textwidth]{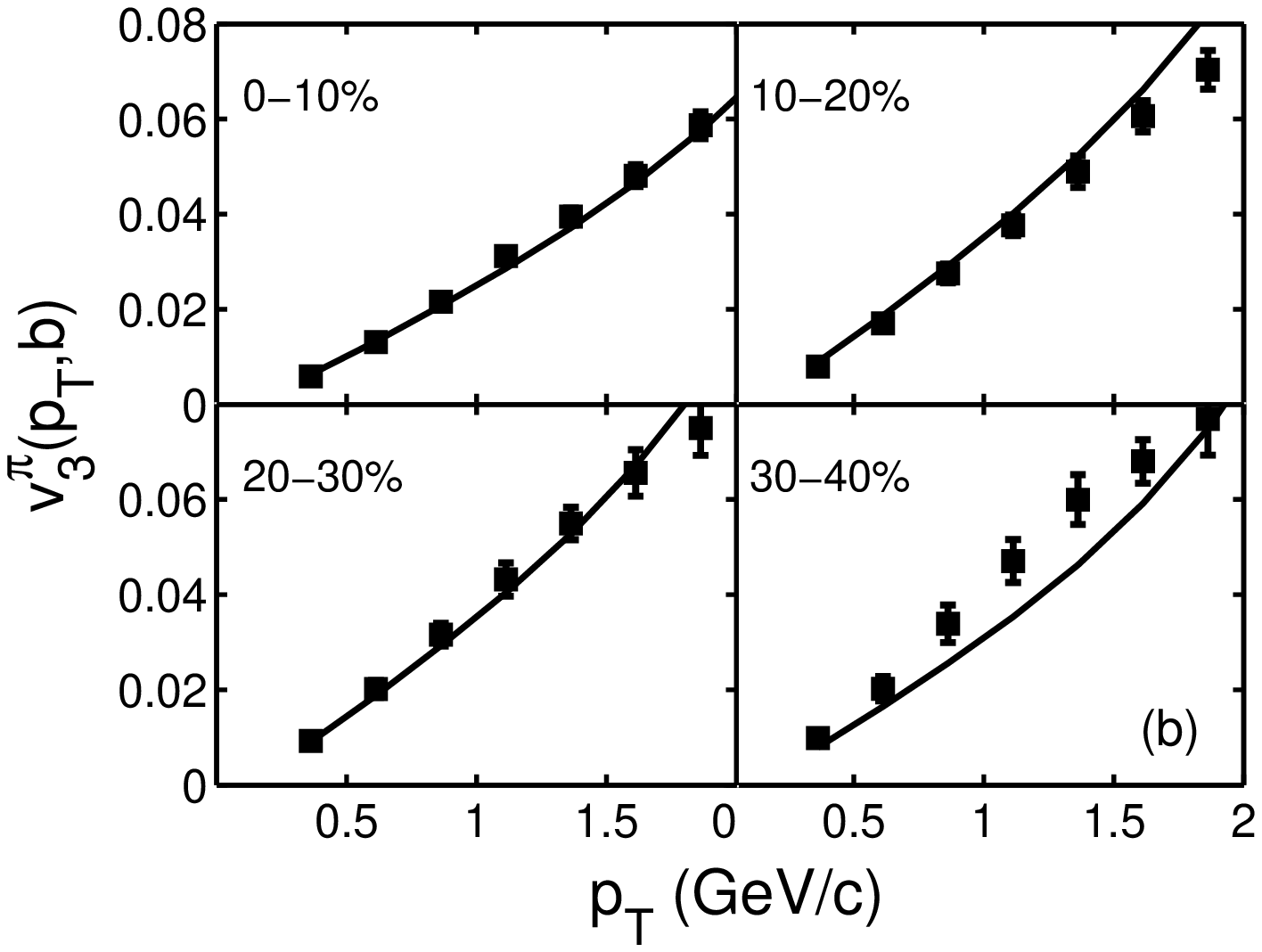}

\includegraphics[width=0.55\textwidth]{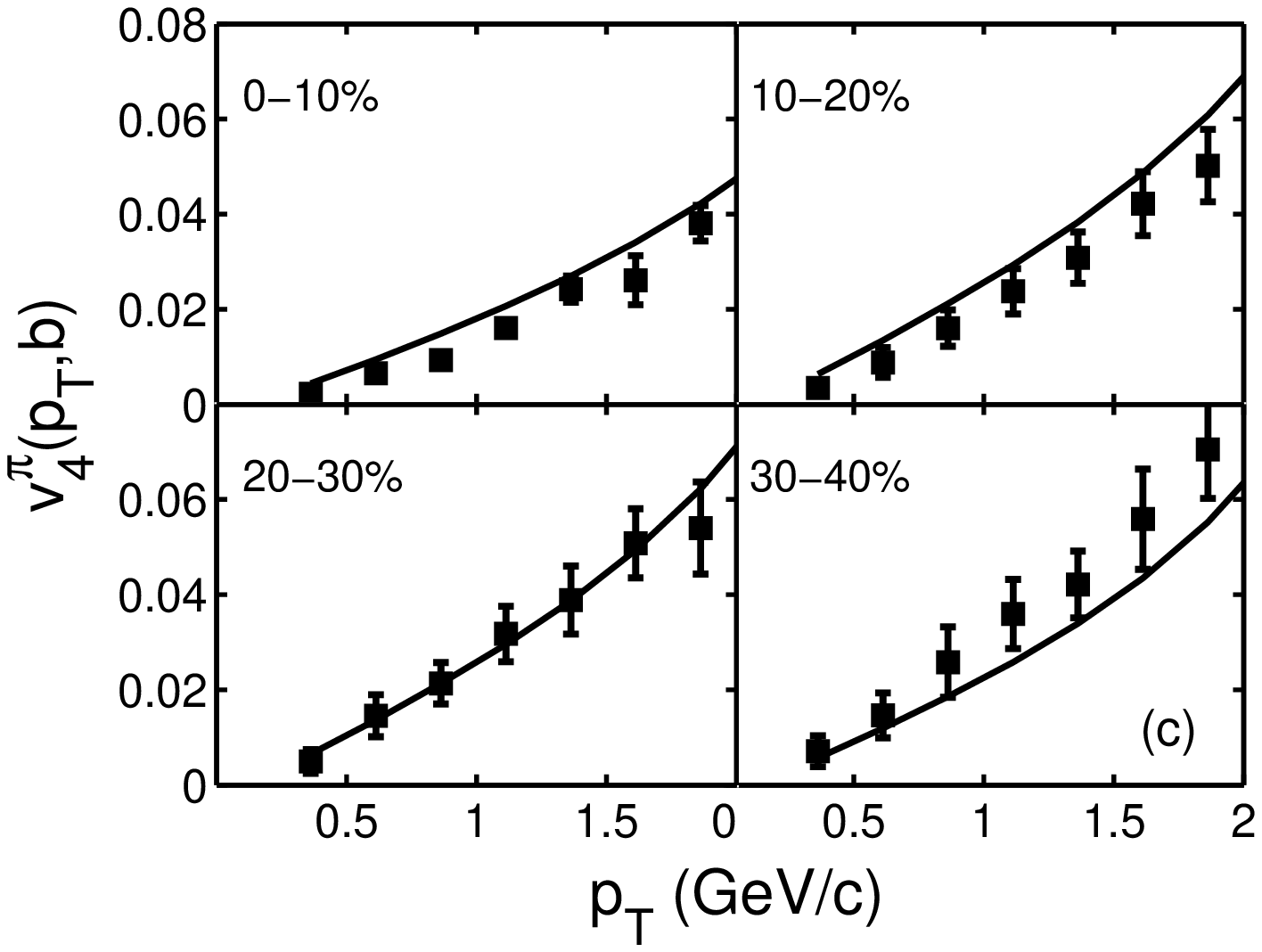}
\caption{$p_T$ dependencies of (a) $v_2^\pi(p_T,b)$, (b) $v_3^\pi(p_T,b)$ and (c) $v_4^\pi(p_T,b)$  for four centralities in each case. Data are from Ref.\ \cite{aa2}. Solid lines are the results of our calculation using $a_n$ given in Eq.\ (\ref{36}).}
\end{figure}

 We note that Fig.\ 7(a) represents an improvement over Fig.\ 5(a) owing to our use of Eq.\ (\ref{30}) instead of (\ref{21}) for $v_2^{\pi}(\pt,b)$. It is clear that the difference is due to the inclusion of the third term in Eq.\ (\ref{10}). However, it is also significant to point out that the change is not large except in the ranges of $\pt$ and $b$. That is, the major features of $v_2^\pi(\pt,b)$ can be well reproduced by considering TT recombination only with the neglect of the TS component.

 The fit of the data is not perfect for $v_3^\pi(\pt,b)$ and $v_4^\pi(\pt,b)$ at 30-40\% centrality. That may be due to larger fluctuations of the minijet contribution at less central collisions, for which more detailed study is called for.

Note that the curvatures of the lines and the data  for $v_2$ are different from those of $v_3$ and $v_4$. To see the origin of that difference, we rewrite Eq.\ (\ref{30}) to reflect the dominance of the numerator of $v_2^\pi(\pt,b)$ by $\bar R^{\pi}(\pt,b)$, and of $v_{3,4}^\pi(\pt,b)$ by $\bar M^{\pi}(\pt,b)$
\bq
v_2^\pi(\pt,b)&\approx& \left<\cos 2\phi\right>_S{\bar R^{\pi}(\pt,b)\over \bar\rho^\pi(\pt,b)} ,  \label{37}  \\
v_{n}^\pi(\pt,b)&=& \left<\cos n\phi\right>_J{\bar M^{\pi}(\pt,b)\over \bar\rho^\pi(\pt,b)}, \qquad n=3,4. \label{38}
\eq
Thus the $\pt$ dependencies of $v_2^\pi(\pt,b)$ and  $v_{3,4}^\pi(\pt,b)$ are dictated by those of $\bar R^{\pi}(\pt,b)/ \bar\rho^\pi(\pt,b)$ and $\bar M^{\pi}(\pt,b)/ \bar\rho^\pi(\pt,b)$, respectively, which are prescribed by the behaviors shown in Fig.\ 6.
$v_2^{\pi}$ is convex upward because, as $\pt$ increases, $\bar M(\pt,b)$ becomes larger so
the increase of $\bar R^{\pi}(\pt,b)/\bar\rho^{\pi}(\pt,b)$ decelerates; in fact, it would decrease as $\pt$ gets above $\pt\sim 3$ GeV/c, a property that has previously been obtained in Ref.\ \cite{chy} because of TS dominance. On the other hand, $v_3^{\pi}$ and $v_4^{\pi}$ are concave upward because $\bar R^{\pi}(\pt,b)$ is much larger than $\bar M^{\pi}(\pt,b)$ around $\pt\sim 1$ GeV/c so $\bar M^{\pi}(\pt,b)/\bar\rho^{\pi}(\pt,b)$ is suppressed at low $\pt$. Eventually, as $\pt$ gets much larger, SS term will become important and turn $v_{3,4}^{\pi}$ over and diminish them.
Since the properties of the three components in Fig.\ 6 are specific results of the recombination model, the appropriate curvatures of the solid lines in Fig.\ 7 in agreement with the data lend  support to our minijet approach to the treatment of azimuthal anisotropy.

Our study here is mainly a demonstration of principle in that minijets are important and can explain all the low-$\pt$ data in the recombination framework. However, it is important to bear in mind that what we have shown is the sufficiency of our approach to reproduce the data, but not necessity. Neither is the hydro approach necessary. Indeed, there is no theoretical treatment that can prove necessity. Nevertheless, it is significant to recognize that various dynamical mechanisms can be responsible for the same phenomenological features of the hadronic observables. By the same token, a combination of those mechanisms may be at play in reality. The base component in our description could possibly be treated by hydrodynamics if thermalization is rapid for the subsystem that is left behind after the emission of semihard parton near surface. There are other related issues concerning the fluctuation of the initial configuration and the variation of the thermalization time for various different eccentricities. Such complications combined with the effects of minijets that we have found here open up a range of possibilities, on which our present treatment may reveal only a restricted view that is opposite to the traditional hydro view. A comprehensive study that includes both components of the mixture is a worthwhile problem for the future. For now, our simple remark is that the common usage of the term "elliptic flow" for $v_2$ is inadequate in generality and tends to be misleading.

\section{conclusion}

We have shown that the major properties of pion and proton production at low $\pt$ can be reproduced in our formulation of hadronization that includes the effects of minijets. The $\pt$ spectra have exponential behavior, $\exp(-\pt/T)$, with a common value of $T$ for both $\pi$ and $p$ that is the same as the $T$ of the thermal partons just before hadronization. Minijets generate azimuthal anisotropy both through energy loss to the medium and in creating shower partons that recombine with the thermal partons. Harmonic analysis of the $\phi$-dependence leads to $v_n(\pt,b)$ that agrees with the data.

We have also shown that the ridge phenomenon is a consequence of minijets. Although our study in this paper is limited to the small-$\eta$ region of inclusive \dis, the ridge component is shown to have a harder $\pt$ spectrum because of the enhancement of the thermal partons. The $\phi$ dependence around a trigger was described in \cite{ch2}, but now we show that when integrated over the trigger direction the ridge component in the inclusive \dis\ generates the quadrupole $v_2(\pt,b)$ with the correct $\pt$ dependence. Thus, $v_2$ and the ridge are tightly related.

Since our treatment is only for the system at late time, we employ no model to carry out the development of the system from early time. The thermal partons are assumed to have an exponential form that is determined by phenomenology. Not following the evolution of the system is not equivalent to an assumption that the system does not expand. It is just that we do not make any assumption concerning the equilibration time or the properties of the fluid. Obviously, we do assume that by the time of hadronization there is local thermalization to justify the use of $T$. The claim  we make is that taking the minijets into account is sufficient to reproduce the measured azimuthal anisotropy. We cannot exclude the validity of hydrodynamical flow, but we do show that the phenomenology that supports the flow dynamics provides the same support for our approach. Thus there are two possible descriptions of the low-$\pt$ process, neither of which can claim exclusive validity. The reality may even be a combination of both.

	While further investigation is needed to determine the extent of the admixture of flow and minijets at RHIC, it is conceivable that in collisions at LHC the density of semihard partons is so high initially that the system has insufficient time for equilibration before the abundant minijets created near the surface dominate the expansion characteristics, even though the higher density of soft gluons speeds up the thermalization process at the core of the medium. If that is so, then one may think of what we have done here as the lower-energy precursor of what needs to be done at higher energies. The study of the $\pt$ spectra at LHC has already shown the importance of minijets through thermal-shower recombination \cite{hz2}. It will therefore be natural to apply the formalism developed here to elucidate  the problem of  azimuthal harmonics measured at LHC.

\section*{Acknowledgment}
This work was supported,  in part,  by the U.\ S.\ Department of Energy under Grant No. DE-FG02-96ER40972 and by the Scientific Research Foundation
for Young Teachers, Sichuan University under No. 2010SCU11090 and
the NSFC of China under Grant No.\ 11147105.

\begin{appendix}
 \section{Centrality Dependence of Hadronic $\pt$ Distributions}

Having obtained the correct centrality dependence of \v\ in Fig.\ 5 of Sec.\ IV, which is totally calculable without free parameters, we consider here the centrality dependence of the inclusive spectra $\bar\rho^h(\pt,b)$. We note that the unknown
 normalization factor ${\cal N}^h$  in Eqs.\ (\ref{6})  never enter into the calculation of \v\ because of cancellation, but for $\bar\rho^h(\pt,b)$ they must be reckoned with. As indicated in  Eq.\ (\ref{7}), ${\cal N}^\pi$ and ${\cal N}^p$
are proportional to $C^2$ and $C^3$, respectively,  due to $q\bar q$ and $qqq$ recombination. The magnitude $C$ of the thermal partons depends on $b$ in a way that cannot be reliably calculated. By phenomenology on the pion spectrum it was previously estimated for $\pt>1.2$ GeV/c \cite{hz}, but that is inadequate for our purpose here; moreover, ${\cal N}^\pi$ and ${\cal N}^p$ have different statistical factors that can depend on $b$ because of resonances. We give here direct parametrizations of the normalization factors in Eq.\ (\ref{7}) in terms of $N_{\rm part}$
\bq
N_0^\pi C^2&=&0.667 N_{\rm part}^{1.05} ,  \label{A1}  \\
N_0^p C^3&=&0.149 N_{\rm part}^{1.18} .  \label{A2}
\eq
Using them in Eqs.\ (\ref{6}) and (\ref{7}) we obtain the curves in Fig.\ 8 (a) pion and (b) proton for three centrality bins. They agree with the data from PHENIX \cite{sa} very well over the range of  $\pt$ shown. In all those curves $T$ is kept fixed at 0.283 GeV, thus reaffirming our point that both pions and protons are produced by the same set of thermal partons despite the apparent differences in the shapes of their $\pt$ dependencies.

\begin{figure}[tbph]
\centering
\hspace*{-7.5cm}
\includegraphics[width=0.5\textwidth]{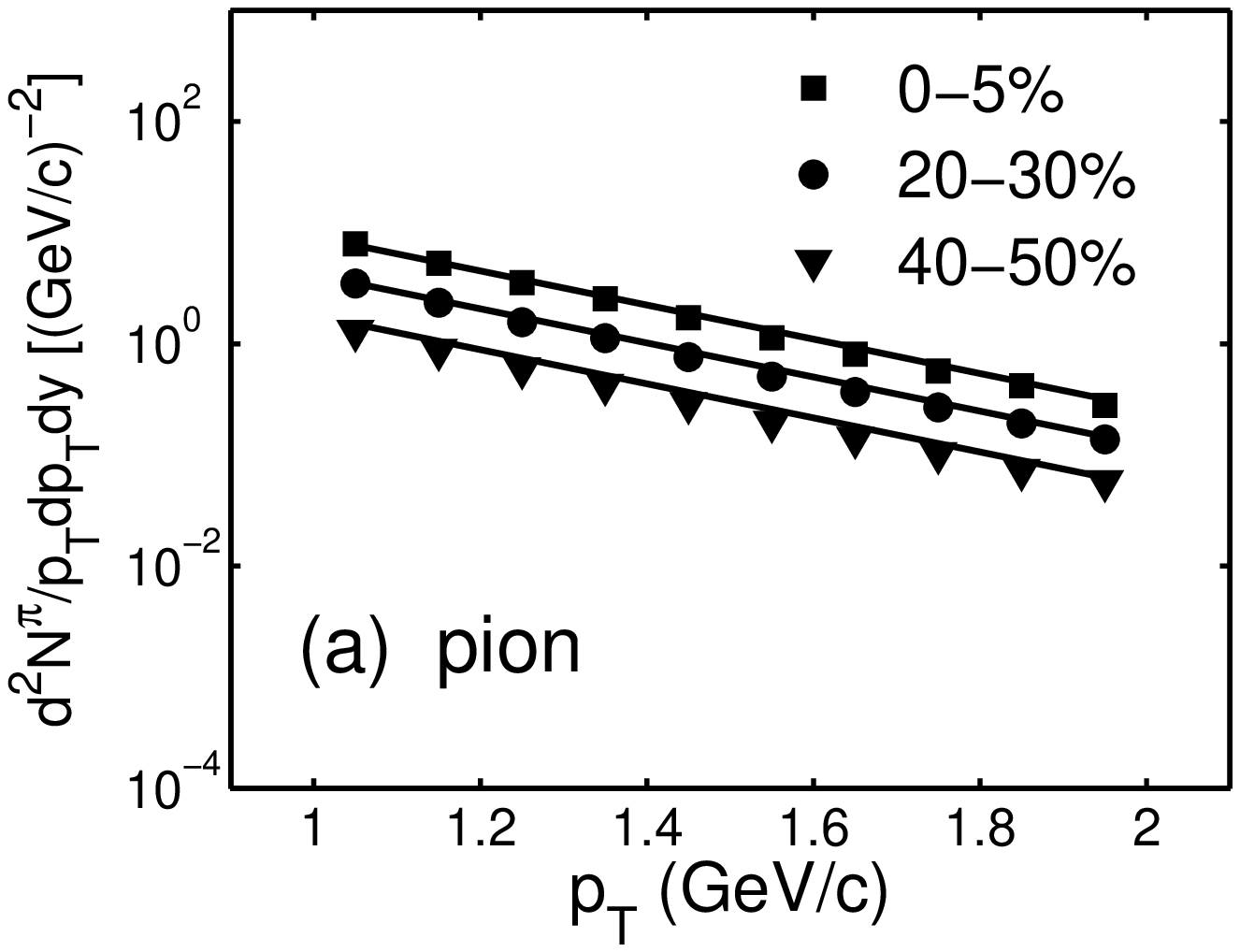}

\centering
\vspace*{-6.2cm}
\hspace*{8.5cm}
\includegraphics[width=0.5\textwidth]{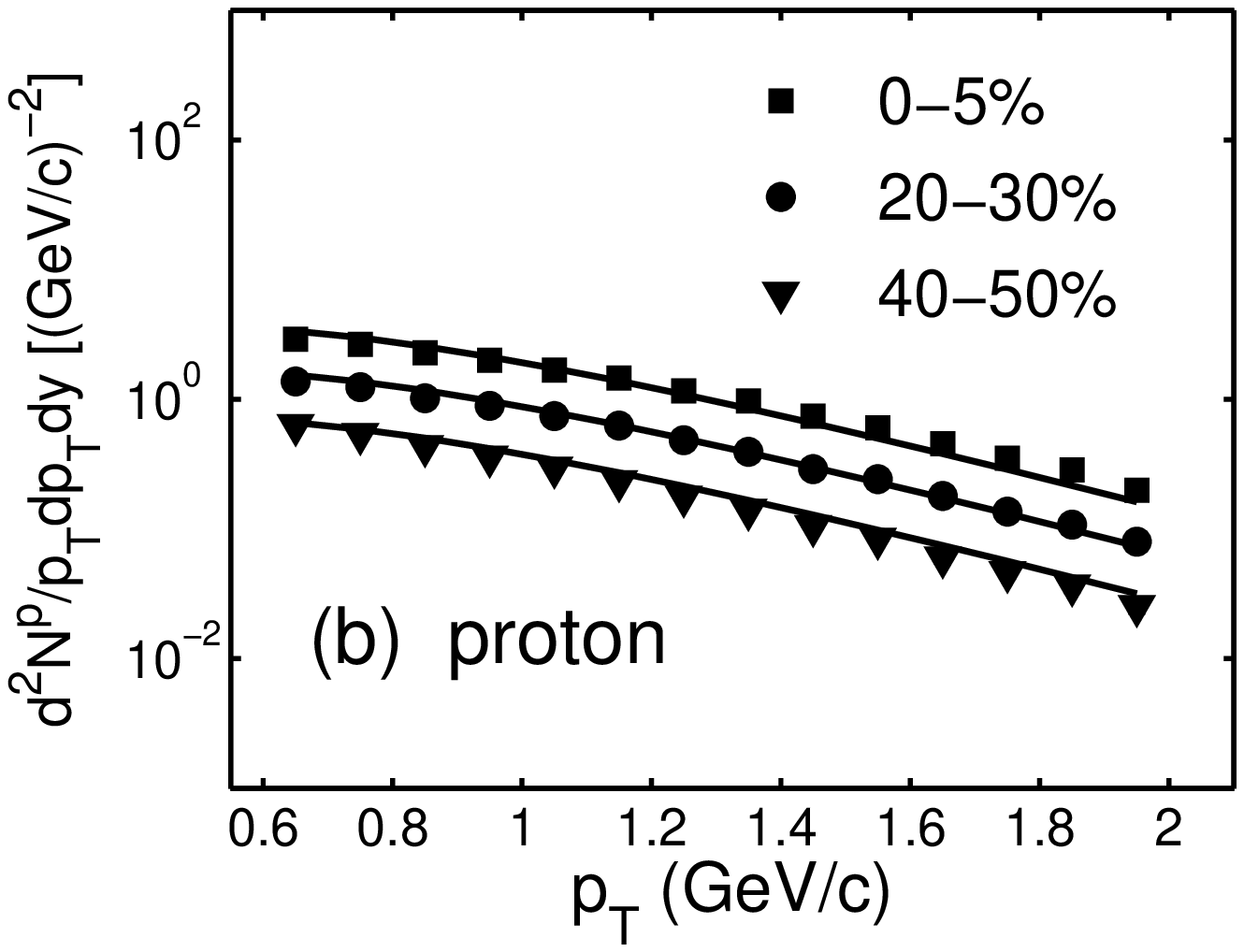}

\caption{Inclusive spectra at three centralities for (a) pion and  (b) proton. The data are from Ref.\ \cite{sa}.}
\end{figure}

\end{appendix}

\end{document}